\documentclass[a4paper]{jpconf}
\usepackage{graphicx}
\begin{document}
\title{An improved nuclear mass formula: WS3}

\author{Ning Wang and Min Liu}

\address{Department of Physics, Guangxi Normal University,
Guilin 541004, P. R. China}

\ead{wangning@gxnu.edu.cn}

\begin{abstract}
We introduce a global nuclear mass formula which is based on the
macroscopic-microscopic method, the Skyrme energy-density functional
and the isospin symmetry in nuclear physics. The rms deviation with
respect to 2149 known nuclear masses falls to 336 keV, and the rms
deviations from 1988 neutron separation energies and $\alpha$-decay
energies of 46 super-heavy nuclei are significantly reduced to 286
and 248 keV, respectively. The predictive power of the mass formula
for describing new measured masses in GSI and those in AME2011 is
excellent. In addition, we introduce an efficient and powerful
systematic method, radial basis function approach, for further
improving the accuracy and predictive power of global nuclear mass
models.

\end{abstract}

\section{Introduction}

It is known that nuclear masses are of great importance for the
study of super-heavy nuclei \cite{WangSHE,Zhang,Lu,Siwek07}, nuclear
symmetry energy \cite{Liu10} and nuclear astrophysics
\cite{Arc,BH,Li11}. For the synthesis of super-heavy nuclei (SHN), a
reliable nuclear mass formula is required for providing some
valuable information on the central position of the "island of
stability", the position of proton drip line and the properties of
SHN at their ground states. In addition, most of masses for nuclei
along the r-process path have not yet been measured. Accurate
predictions for the masses of these extremely neutron-rich nuclei
play a key role for the study of nuclear astrophysics. Available
nuclear mass formulas include some global and local formulas. For
the global formulas, the model parameters are usually determined by
all measured masses and the masses of almost all bound nuclei can be
calculated.  Some global nuclear mass models have been successfully
established, such as the finite range droplet model (FRDM)
\cite{Moll95} which is based on the macroscopic-microscopic method,
the Hartree-Fock-Bogliubov (HFB) model \cite{HFB17}, the Duflo-Zuker
formula \cite{DZ10,DZ28} and the Weizs\" acker-Skyrme formula
\cite{Wang,Wang10,Wang11}. These models can reproduce the measured
masses with accuracy at the level of 600 to 300 keV. The local mass
formulas are generally based on algebraic or systematic approaches.
They predict the masses of unknown nuclei from the masses of known
neighboring nuclei, such as the Garvey-Kelson relations \cite{GK},
the isobaric multiplet mass equation, the residual proton-neutron
interactions \cite{Zhao,Zhao12} and the image reconstruction
technique (like the CLEAN algorithm \cite{Mora}  and the radial
basis function method \cite{RBF}). The main difficult of the local
mass formulas is that the model errors rapidly increase for nuclei
far from the measured nuclei. For neutron drip line nuclei and
super-heavy nuclei, the differences between the calculated masses
from these different models are quite large. It is therefore
necessary to check the reliability and predictive power of the
nuclear mass formulas.

\section{Weizs\" acker-Skyrme mass formula}

In Ref. \cite{Wang11} an improved nuclear mass formula, Weizs\"
acker-Skyrme mass formula, was proposed. The total energy of a
nucleus is written as a sum of the liquid-drop energy, the
Strutinsky shell correction $\Delta E$ and the residual correction
$\Delta_{\rm res}$,
\begin{eqnarray}
E (A,Z,\beta)=E_{\rm LD}(A,Z) \prod_{k \ge 2} \left (1+b_k \beta_k^2
\right )+\Delta E (A,Z,\beta) + \Delta_{\rm res}.
\end{eqnarray}
The liquid-drop energy of a spherical nucleus $E_{\rm LD}(A,Z)$ is
described by a modified Bethe-Weizs\"acker mass formula,
\begin{eqnarray}
E_{\rm LD}(A,Z)=a_{v} A + a_{s} A^{2/3}+ E_C + a_{\rm sym} I^2 A +
a_{\rm pair}  A^{-1/3}\delta_{np} + \Delta_W
\end{eqnarray}
with the isospin asymmetry $I=(N-Z)/A$ and the Coulomb energy,
\begin{eqnarray}
E_C=a_c \frac{Z^2}{A^{1/3}} \left ( 1- 0.76 Z^{-2/3} \right).
\end{eqnarray}

In the original Bethe-Weizs\"acker formula, the symmetry energy
coefficient $a_{\rm sym}$ is simply written as a constant. To
consider the surface-symmetry energy of finite nuclei, the
mass-dependence of the $a_{\rm sym}$ was proposed in Refs.
\cite{Myers,Dani}.  To check the mass-dependence of $a_{\rm sym}$,
we have studied the symmetry energy coefficients of finite nuclei
based on the experimental data for masses \cite{WangWig}. The
liquid-drop energy which is a function of mass number $A$ and charge
number $Z$ can also be expressed as a function of $A$ and the
isospin asymmetry $I$. By performing a partial derivative of the
liquid-drop energy per particle with respect to  $I$, the symmetry
energy coefficients of finite nuclei can be extracted. In addition
to the mass-dependence of $a_{\rm sym}$, the isospin dependence of
the symmetry energy coefficients of finite nuclei can also be
clearly observed. To consider the isospin dependence of $a_{\rm
sym}$, we proposed a new form for the symmetry energy coefficient
\begin{eqnarray}
 a_{\rm sym}=c_{\rm sym}\left [1-\frac{\kappa}{A^{1/3}}+ \xi  \frac{2-|I|}{ 2+|I|A}  \right
 ],
\end{eqnarray}
with which the mass- and isospin-dependence of $a_{\rm sym}$ can be
reproduced reasonably well. The $|I|$ term in  $a_{\rm sym}$
represents the traditional Wigner effect \cite{WangWig}. With the
proposed form for $a_{\rm sym}$, we found that the accuracy of the
liquid-drop formula can be improved by $\sim 5\%$ and the value of
$c_{\rm sym}$ which represents the nuclear symmetry energy at normal
density goes up to about 30 MeV (very close to the extracted value
from other approaches). Furthermore, it was found that the model
parameters in the liquid-drop formula with the proposed form of
$a_{\rm sym}$ are quite stable for different mass region
\cite{Hirsch}.

The $a_{\rm pair}$ term empirically describes the pairing effect
with
\begin{eqnarray}
\delta_{np}= \left\{
\begin{array} {r@{\quad:\quad}l}
  2 - |I|  &   N {\rm ~and~} Z {\rm ~even }    \\
      |I|  &   N {\rm ~and~} Z {\rm ~ odd }    \\
  1 - |I|  &   N {\rm ~even,~} Z {\rm ~odd,~ } {\rm ~and~ } N>Z   \\
  1 - |I|  &   N {\rm ~odd,~} Z {\rm ~even,~ } {\rm ~and~ } N<Z   \\
  1             &   N {\rm ~even,~} Z {\rm ~odd,~ } {\rm ~and~ } N<Z   \\
  1             &   N {\rm ~odd,~} Z {\rm ~even,~ } {\rm ~and~ } N>Z   \\
\end{array} \right.
\end{eqnarray}
The influence of nuclear deformations on the liquid-drop energy is
considered based on a parabolic approximation. We studied the
variation of the energy of a nucleus as a function of $\beta_2$ and
$\beta_4$ by using the Skyrme energy-density functional plus the
extended Thomas-Fermi approach \cite{Liu06}, and found that the
parabolic approximation is applicable for small deformations. The
terms with $b_k$ in Eq.(1) describe the contribution of nuclear
deformation (including $\beta_2$, $\beta_4$ and $\beta_6$) to the
macroscopic energy. Mass dependence of the curvatures $b_k$  is
written as \cite{Wang},
\begin{eqnarray}
b_k=\left ( \frac{k}{2} \right ) g_1A^{1/3}+\left ( \frac{k}{2}
\right )^2 g_2 A^{-1/3},
\end{eqnarray}
according to the Skyrme energy-density functional, which greatly
reduces the computation time for the calculation of deformed nuclei.

The microscopic shell correction
\begin{eqnarray}
\Delta E=c_1 E_{\rm sh} + |I| E_{\rm sh}^{\prime}
\end{eqnarray}
is obtained with the traditional Strutinsky procedure  by setting
the order $p=6$ of the Gauss-Hermite polynomials and the smoothing
parameter $\gamma=1.2\hbar\omega_0$ with $\hbar\omega_0=41 A^{-1/3}$
MeV. $E_{\rm sh}$ and $E_{\rm sh}^{\prime}$ denote the shell energy
of a nucleus and of its mirror nucleus, respectively. The $|I|$ term
in $\Delta E$ is to take into account the mirror nuclei constraint
\cite{Wang10} from the isospin symmetry, with which the accuracy of
the mass model can be significantly improved by $10\%$ without
introducing any new model parameters. The charge-symmetry and
charge-independence of nuclear force implies the energies of a pair
of mirror nuclei should be close to each other if removing the
Coulomb term. Combining the macro-micro method, one finds that the
difference of the shell correction for mirror nuclei should be close
to each other. The experimental data also support this point. From
the traditional Strutinsky shell correction calculations, we note
that the difference of the shell correction reaches a few MeV for
some nuclei. The new expression for the shell correction in which
the shell energy of the mirror nucleus is also involved, is proposed
to consider the mirror effect.

In the calculations of the shell corrections, the single-particle
levels are obtained under an axially deformed Woods-Saxon potential
\cite{Cwoik}. Simultaneously, the isospin-dependent spin-orbit
strength is adopted based on the Skyrme energy-density functional,
\begin{eqnarray}
\lambda= \lambda_0 \left ( 1+\frac{N_i}{A} \right )
\end{eqnarray}
with $N_i=Z$ for protons and $N_i=N$ for neutrons, which strongly
affects the shell structure of neutron-rich nuclei and super-heavy
nuclei.

In this formula, we also consider the Wigner effect of heavy nuclei
coming from the approximate symmetry between valence protons and
neutrons. We found that some heavy doubly magic nuclei, such as
$^{132}$Sn, $^{208}$Pb and $^{270}$Hs, lie along a straight line
$N=1.37Z+13.5$. The shell corrections of nuclei are approximately
symmetric along this straight line. The Wigner effect of heavy
nuclei causes the nuclei along this line are more bound. The term
$\Delta_W$ is to consider this effect and reduces the rms error by
$\sim 5\%$ (see Ref.\cite{Wang11} for details).

Finally, we would like to introduce a very efficient and powerful
systematic correction to the global nuclear mass formulas - radial
basis function (RBF) correction \cite{RBF}. The RBF approach is a
prominent global interpolation and extrapolation scheme for
scattered data fitting. It is widely used in surface reconstruction.
Based on a global nuclear mass formula and known masses, the
differences between model calculation and experimental data are
analyzed. The aim of the RBF approach is to find a smooth function
$S(N,Z)$ to describe the differences  $M_{\rm exp}-M_{\rm th}$. Once
the reconstructed function $S$ is obtained, the revised masses for
unmeasured nuclei are given by $M^{\rm RBF}_{\rm th}=M_{\rm th}+S$.
Here, $M_{\rm th}$ denotes the calculated mass from a global nuclear
mass model. We find that the RBF correction can improve the accuracy
of a global mass formula by $\sim 10\%$ to $40\%$ without
introducing any new model parameters.

\section{Results}

\begin{center}
\begin{table}[h]
\centering \caption{ rms deviations  between data AME2003
\cite{Audi} and predictions of some models (in keV). The line
$\sigma (M)$ refers to all the 2149 measured masses, the line
$\sigma (S_n)$ to the 1988 measured neutron separation energies
$S_n$, the line $\sigma (Q_\alpha)$ to the $\alpha$-decay energies
of 46 super-heavy nuclei. }

\centering
\begin{tabular}{cccccccccc}
\br
       &  FRDM  & HFB-17 & DZ10 & WS & WS*& DZ31 & DZ28  & WS3 \\
\mr
 $\sigma  (M)$        & $656$  & $581$ & $561$ & $516$ & $441$ & $362$ & $360$ &  $336 $\\
 $\sigma  (S_n)$      & $399$  & $506$ & $342$ & $346$ & $332$ & $299$ & $306$ &  $286 $\\
 $\sigma  (Q_\alpha)$ & $566$  & $-$   & $916$ & $284$ & $263$ & $1052$& $936$ &  $248 $\\
 Reference            & \cite{Moll95}  & \cite{HFB17} & \cite{DZ10}
 & \cite{Wang} & \cite{Wang10} & \cite{DZ10} & \cite{DZ28} & \cite{Wang11} \\

\br
\end{tabular}
\end{table}
\end{center}

\begin{figure}
\includegraphics[angle=-0,width= 1.0\textwidth]{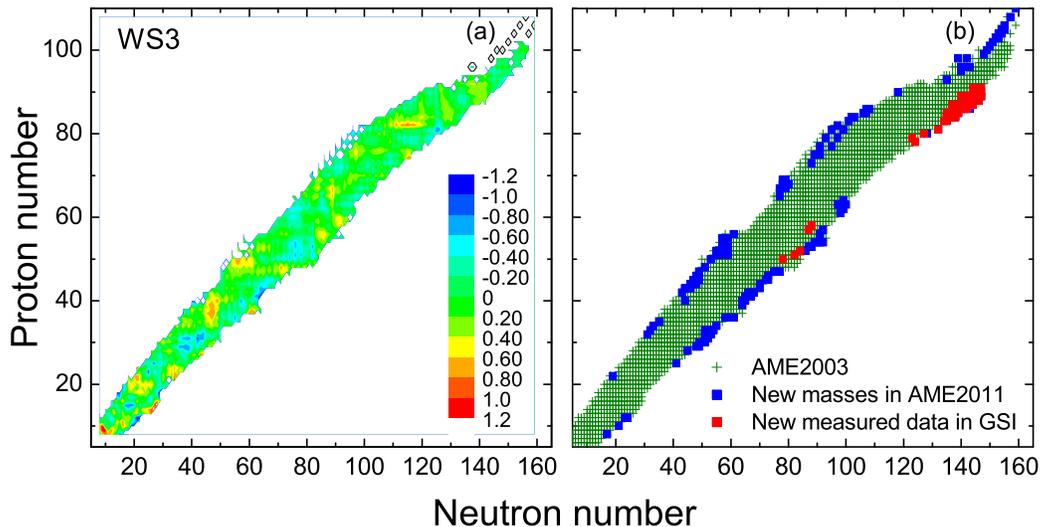}
 \caption{(Color online) (a) Difference between the calculated masses with
 the WS3 formula \cite{Wang11} and the experimental data AME2003. (b) Positions of new measured masses.
 The blue and red squares denote the new measured nuclei in AME2011 \cite{AME2011} and those in GSI \cite{GSImass}, respectively.  }
\end{figure}

The rms deviations between the 2149 experimental masses AME2003
\cite{Audi} and predictions of some models are calculated,
\begin{eqnarray}
  \sigma(M)=\left [\frac{1}{m}\sum \left (M_{\rm exp}^{(i)}-M_{\rm
th}^{(i)}\right )^2 \right ] ^{1/2} .
\end{eqnarray}
We show in Table 1 the calculated rms deviations $\sigma (M)$ (in
keV). $\sigma (S_n)$ denotes the rms deviation to the 1988 measured
neutron separation energies $S_n$. $\sigma (Q_\alpha)$ denotes the
rms deviation with respect to the $\alpha$-decay energies of 46
super-heavy nuclei ($Z\ge 106$) \cite{Wang10}.  FRDM and HFB-17
denote the finite-range droplet model \cite{Moll95} and the latest
Hartree-Fock-Bogoliubov (HFB) model with the improved Skyrme
energy-density functional \cite{HFB17}, respectively. DZ10, DZ28 and
DZ31 denote the Duflo-Zuker mass models with 10, 28 and 31
parameters, respectively \cite{DZ10,DZ28}. The rms deviation from
the 2149 masses with WS3 is remarkably reduced to 336 keV, much
smaller than the results from the  FRDM and the latest HFB-17
calculations, even lower than that achieved with the best of the
Duflo-Zuker models. The rms deviation with respect to the 1988
neutron separation energies is reduced to 286 keV, and the rms
deviation to the $\alpha$-decay energies of 46 super-heavy nuclei is
reduced to 248 keV, much smaller than the result of Duflo-Zuker
formula (which is about one MeV). Fig. 1 (a) shows the difference
between the experimental data AME2003 and the calculated masses with
the WS3 formula \cite{Wang11} ($M_{\rm exp}-M_{\rm th}$). One sees
that most of nuclei can be described remarkably well and the
deviations for some semi-magic nuclei are slightly large.

\begin{center}
\begin{table}[h]
\caption{ rms deviations with respect to 154 new masses in AME2011
\cite{AME2011}
 and those to 53 new measured masses in GSI \cite{GSImass} from four
 mass models (in keV). }

\centering
\begin{tabular}{ccccc}
\br
                & FRDM & HFB17 & DZ28 & WS3 \\
\mr
154 new masses in AME2011 & 723  & 693   & 622  & 433 \\
53 new  masses in GSI     & 600  & 582   & 504  & 360 \\

\mr
\end{tabular}
\end{table}
\end{center}

\begin{center}
\begin{table}[h]
\caption{The same as Table 2, but the radial basis function (RBF)
corrections are involved. }

\centering
\begin{tabular}{ccccc}
 \br
                & FRDM & HFB17 & DZ28 & WS3 \\
\mr
154 new masses in AME2011 & 475  & 596   & 442  & 397 \\
53 new  masses in GSI     & 193  & 250   & 287  & 229 \\

 \mr
\end{tabular}
\end{table}
\end{center}

Here, we also check the predictive power of different mass formulas
for description of new masses. The crosses in Fig. 1(b) denote the
available masses in AME2003. After 2003, 154 new measured data have
been collected by Audi and Wang (blue squares) in the interim
AME2011 \cite{AME2011} and 53 new masses are measured in GSI
\cite{GSImass} (red squares). The rms deviations with respect to
these new data are listed in Table 2. For the new masses in AME2011,
the corresponding results of FRDM and HFB are about 700 keV and the
result of DZ28 model is more than 600 keV. The result of WS3 is only
433 keV. For the 53 new data in GSI, the result from WS3 is also the
smallest one (only 360 keV).

\begin{figure}
\includegraphics[angle=-0,width= 0.8\textwidth]{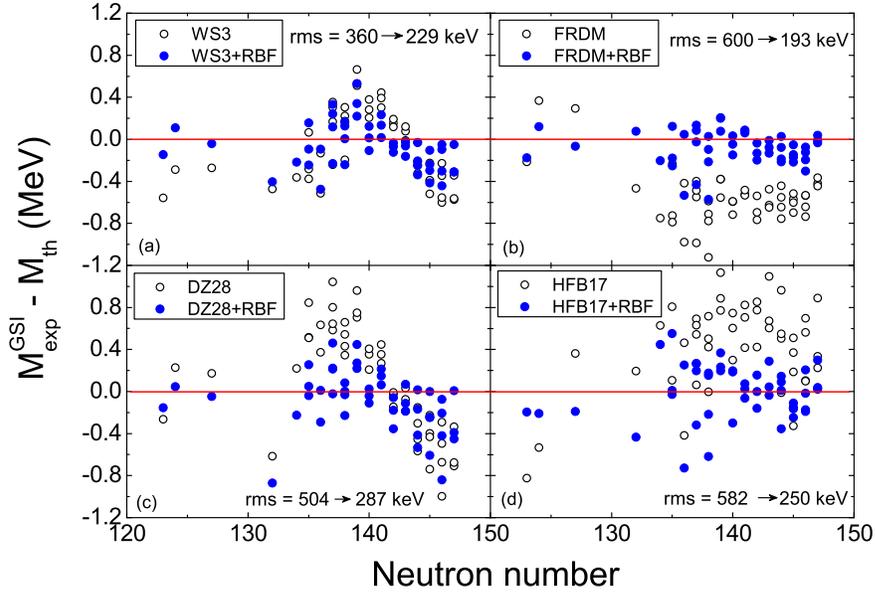}
 \caption{(Color online) Difference between model predictions and the experimental data for the new measured masses in GSI.   }
\end{figure}

\begin{figure}
\includegraphics[angle=-0,width= 0.85\textwidth]{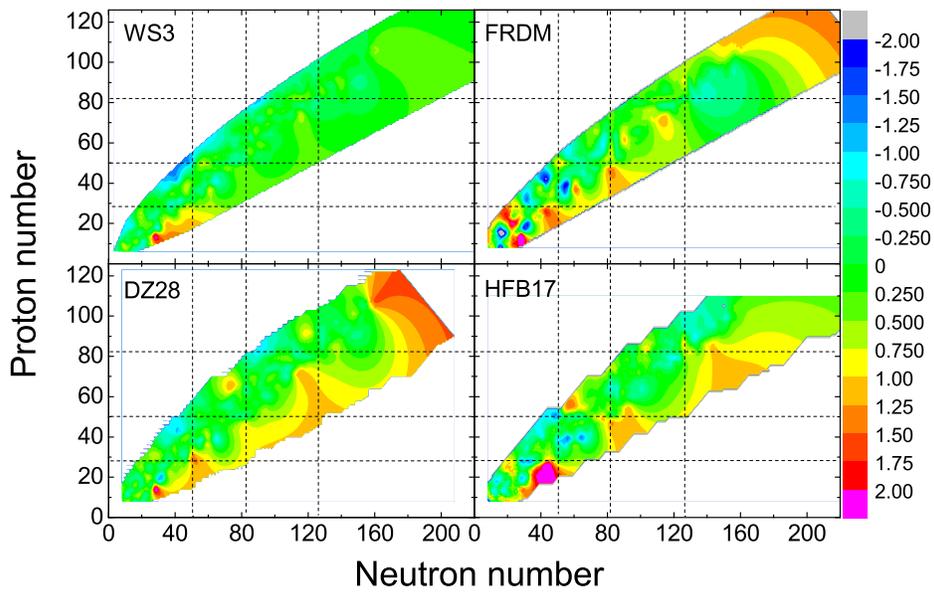}
 \caption{(Color online) Radial basis function corrections for different global nuclear mass models.
 The dashed lines show the magic numbers.   }
\end{figure}

Combining the radial basis function correction, we find that the
accuracy of global mass formulas can be significantly improved. With
the RBF corrections (see Table 3), the rms error to the new data in
AME2011 is reduced from 723 keV to 475 keV and the rms error to the
new data in GSI is reduced from 600 keV to about 200 keV based on
the FRDM calculations. For other models, the improvement is also
remarkable. Fig. 2 shows the difference between different model
predictions and experimental data for the new masses measured in GSI
\cite{GSImass}. The solid circles denote the results when the RBF
corrections are involved. For these global nuclear mass models,
especially the FRDM and HFB17 models, the RBF approach plays an
important role for improving the systematic errors. Here, the
measured masses in AME2003 are adopted for training the RBF [to
obtain the function $S(N,Z)$] based on the leave-one-out
cross-validation method. Simultaneously, the Garvey-Kelson relation
\cite{GK}, which contains 12 estimates for a nucleus with the
corresponding values of its 21 neighbors, is also adopted for
further improving the smoothness of the function $S(N,Z)$. For
unmeasured nuclei, the masses are predicted by using $M^{\rm
RBF}_{\rm th}=M_{\rm th}+S$ with the calculated mass $M_{\rm th}$
from a global nuclear mass model. Fig. 3 shows the RBF corrections
for different global nuclear mass models. For the super-heavy nuclei
region, the RBF corrections are large for the FRDM and DZ28 model.
For the DZ28 model, one can see from Table 1 that the rms deviation
with respect to the known masses is only 360 keV, however, the rms
deviation $\sigma(Q_\alpha)$ with respect to the $\alpha$-decay
energies of 46 super-heavy nuclei goes up to 936 keV, which also
implies that the systematic errors for the masses of super-heavy
nuclei  could be large  in this model.

\begin{figure}
\includegraphics[angle=-0,width= 0.85\textwidth]{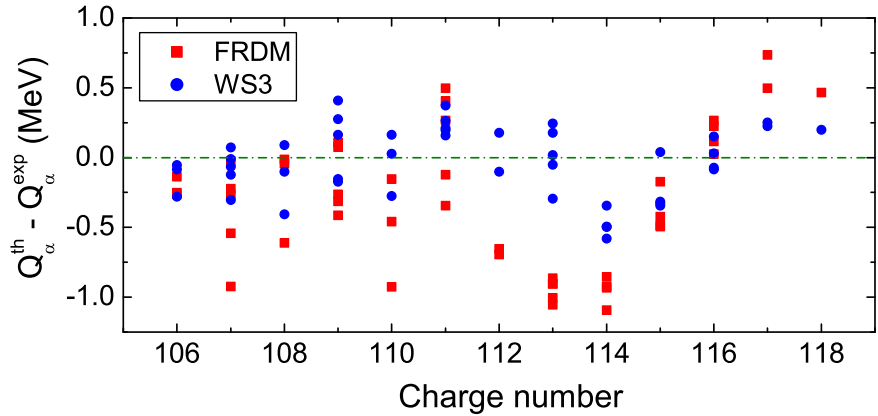}
 \caption{(Color online) Difference between model predictions and the experimental data for the $\alpha$-decay energies of super-heavy nuclei.
 The squares and circles denote the results of FRDM and WS3, respectively.  }
\end{figure}

\begin{figure}[h]
\begin{minipage}{18pc}
\includegraphics[width=18 pc]{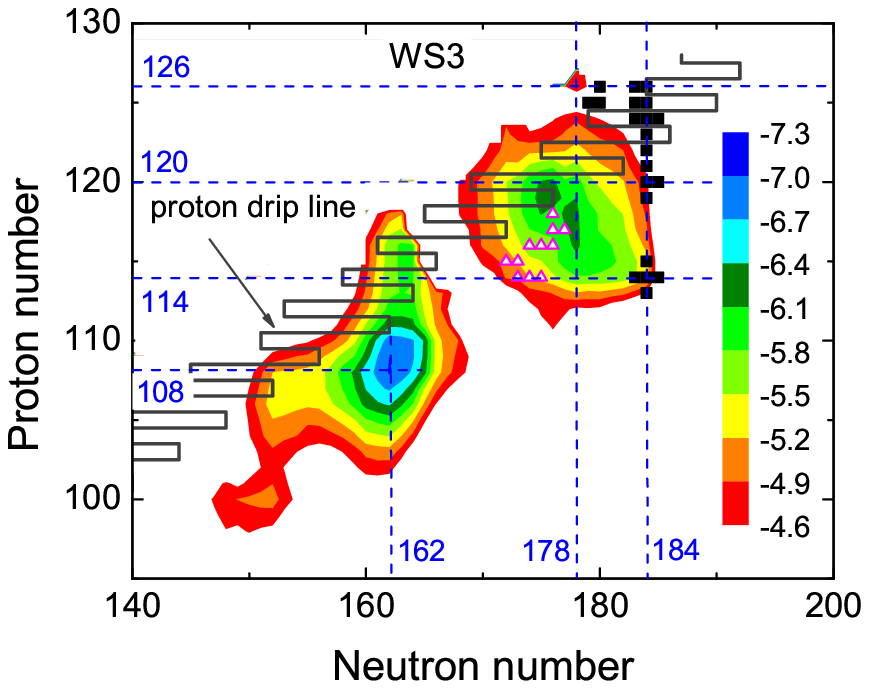}
\caption{\label{label}  Shell corrections of nuclei in super-heavy
region from WS3 calculations. The black squares denote the nuclei
with spherical shapes and the triangles denote the synthesized SHN
in Dubna. The dark gray zigzag line denotes the calculated proton
drip line. The dashed lines show the possible magic numbers.}
\end{minipage}\hspace{2pc}%
\begin{minipage}{18pc}
\includegraphics[width=18pc]{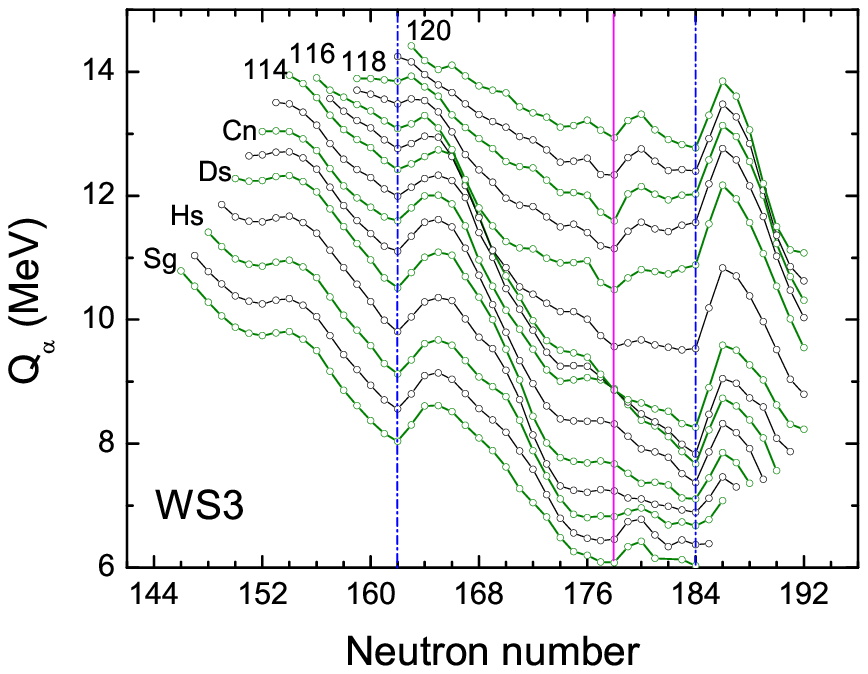}
\caption{\label{label} Predicted (ground state to ground state)
$\alpha$-decay energies of super-heavy nuclei. The thick and thin
curves denote the results for even-$Z$ and odd-$Z$ nuclei,
respectively. The two dashed lines show neutron numbers $N=162$ and
$184$, and the solid line show the position of $N=178$. \\}
\end{minipage}
\end{figure}

It is known that the $\alpha$-decay energies of super-heavy nuclei
have been measured with a high precision, which provides us with
useful data for testing mass models. Fig. 4 shows the difference
between model predictions and the experimental data for the
$\alpha$-decay energies of SHN.  Comparing with the FRDM, the
$\alpha$-decay energies of the super-heavy nuclei are much better
reproduced with the WS3 model. The difference is within $\Delta
Q_\alpha=\pm 0.5$ MeV for SHN with the WS3 formula. The
corresponding result from the FRDM is $ \pm 1.0$ MeV. Here, we would
like to emphasize that the description for the $\alpha$-decay
energies of SHN is also a useful tool to test the predictive power
of the mass models because the measured $\alpha$-decay energies are
not involved in the fit for the model parameters.

Based on the WS3 formula, the shell corrections and the surface of
the $\alpha$-decay energies of super-heavy nuclei have been studied
simultaneously. In Fig. 5, we show the calculated shell corrections
$\Delta E$ of nuclei. There are two islands in the super-heavy mass
region. One is located at $N=162$, $Z=108$ and the other is located
around $N=178$, $Z=118$. The triangles denote the synthesized SHN in
Dubna through fusion reactions by using $^{48}$Ca bombarding on
actinide targets. The calculated deformations of nuclei demonstrate
that the nuclei with $N=184$ are (nearly) spherical in shape.
However, the maximum of the shell corrections occurs at around
$N=178$ instead of $N=184$. The shell corrections (in absolute
value) for nuclei with $Z=126$ are smaller than that for nucleus
($N=178$, $Z=118$) by more than one MeV. Furthermore, nuclei with
$Z=126$ and $N \le 184$ locate beyond the calculated proton drip
line, which indicates that the probability to synthesize nuclei with
$Z=126$ would be much smaller than that of produced super-heavy
nuclei already. Fig. 6 shows the predicted $\alpha$-decay energies
of super-heavy nuclei with the WS3 model. The two valleys at $N=162$
and 184 can be clearly observed. One can also see the valley at
$N=178$. For SHN with $Z=116 \sim 118$, the $\alpha$-decay energies
for nuclei with $N=178$ are smaller than those for nuclei with
$N=184$.

\section{Summary}

In this talk, we introduced a global nuclear mass formula, Weizs\"
acker-Skyrme mass formula, which is based on the Skyrme
energy-density functional, the macroscopic-microscopic method and
the isospin symmetry in nuclear physics. The rms deviations from
2149 measured masses and 1988 neutron separation energies are
significantly reduced to 336 and 286 keV, respectively. As a test of
the extrapolation of the mass model, the $\alpha$-decay energies of
46 super-heavy nuclei have been systematically studied. The rms
deviation with the proposed model falls to 248 keV, much smaller
than the result 936 keV from the DZ28 model. Shell effect and
isospin effect, especially isospin symmetry play a crucial role for
improving the nuclear mass formula. For further testing the
predictive power of the nuclear mass models, 154 new measured data
in AME2011 and 53 new masses measured in GSI have been studied. The
Weizs\" acker-Skyrme mass formula can reproduce these new data
remarkably well. We also introduced an efficient and powerful
systematic method, radial basis function (RBF) approach, for further
improving the accuracy and predictive power of global nuclear mass
models. With the proposed mass formula, the shell corrections and
$\alpha$-decay energies of super-heavy nuclei have been studied
simultaneously. The calculated shell corrections and $\alpha$-decay
energies of super-heavy nuclei imply that the shell gap at $N = 178$
also influences the central position of the "island of stability"
for SHN.

\section{Acknowledgments} This work was supported by
National Natural Science Foundation of China, Nos 11275052,
11005022, 10875031 and 10847004. The obtained mass table with the
formula WS3 is available at http://www.imqmd.com/wangning/WS3.6.txt.
The nuclear mass tables with the radial basis function corrections
are available at http://www.imqmd.com/wangning/RBFmasses.txt.

\end{document}